\documentclass{ws-ijmpa}
\usepackage{graphicx}
\usepackage{amsfonts}
\usepackage{amssymb}%

\providecommand{\Journal}[4] {#1 {\bf #2}, #3 (#4)}
\providecommand{\PLB}{Phys. Lett. B} %
\providecommand{\PRL}{Phys. Rev. Lett.} %
\providecommand{\PRD}{Phys. Rev. D}
\providecommand{\NT}{Nature}
\providecommand{\SCI}{Science}
\providecommand{\PR}{Phys. Rev.} %
\providecommand{\JMP}{J. Math. Phys.}%
\providecommand{\AoP}{Annals of Physics}%
\providecommand{\AP}{Ann.Phys.}%
\providecommand{\ZET}{Zh.\ EksP.\ Teor.\ Fiz.\ Pis'ma Red.}%
\providecommand{\JETPl} {JETP Lett.}%
\providecommand{\IB}{ibid.}%

\newcommand{\be}{\begin{equation}}
\newcommand{\ee}{\end{equation}}
\newcommand{\bea}{\begin{eqnarray}}
\newcommand{\eea}{\end{eqnarray}}

\begin{document}
%\markboth{Z.~Xiao \& B.-Q.~Ma} {Instructions for Typing Manuscripts
%(Lorentz violation dispersion relation and its application)}
\catchline{}{}{}{}{}
%
%%%%%%%%%%%%%%%%%%%%%%%%%%%%%%%%%%%%%%%%%%%%%%%%%%%%%%%%%%%%%%%%%%%%
\title{Lorentz violation dispersion relation and its application}

\author{Zhi Xiao}
\address{School of Physics and State Key Laboratory of Nuclear
Physics and Technology, Peking University, Beijing 100871, China}
\author{Bo-Qiang Ma\footnote{Corresponding author. Email address: \texttt{mabq@phy.pku.edu.cn}}}
%\affiliation{School of Physics and State Key Laboratory of Nuclear
%Physics and Technology, Peking University, Beijing 100871, China}
\address{School of Physics and State Key Laboratory of Nuclear
Physics and Technology, Peking University, Beijing 100871, China}

\maketitle

\begin{abstract}
We derive a modified dispersion relation (MDR) in the Lorentz
violation extension of quantum electrodynamics (QED) sector in the
standard model extension (SME) framework. Based on the extended
Dirac equation and corresponding MDR, we observe the resemblance of
the Lorentz violation coupling with spin-gravity coupling. We also
develop a neutrino oscillation mechanism induced by the presence of
nondiagonal terms of Lorentz violation couplings in 2-flavor space
in a 2-spinor formalism by explicitly assuming neutrinos to be
Marjorana fermions. We also obtain a much stringent bound
($\backsim10^{-25}$) on one of the Lorentz violation parameters by
applying MDR to the ultrahigh energy cosmic ray (UHECR) problem.

\end{abstract}

%\pacs{11.30.Cp, 11.30.Er, 14.60.Pq, 98.70.Sa} \maketitle
\ccode{PACS numbers: 11.30.Cp, 11.30.Er, 14.60.Pq, 98.70.Sa}

%\maketitle
\section{Introduction\label{sec:1}}
In the development of physics, symmetry principle is a powerful tool
in the construction and interpretation of physical laws of nature.
Various efforts have been dedicated to the searches of new symmetry
principle beyond standard model (SM) gauge symmetry and ordinary
Lorentz symmetry, such as $SU(5)$ and $SO(10)$ in grand unified
theory (GUT), or $SO(32)$ and $E(8)\bigotimes E(8)$ in string
theory. Aside from these gauge symmetries, ordinary Lorentz symmetry
is also extended to $SO(9,1)$ in string theory or $SO(10,1)$ in
M-theory.

On the other hand, symmetry principle is not implemented trivially
in nature. The discovery of non-conservation of parity in
1957\cite{Parity} makes people to realize that some sacred symmetry
may be only a good approximation. And the later discovery of
electroweak theory teaches us that symmetry could be hidden in
vacuum condensation, in other words, it is realized through
spontaneous symmetry breaking mechanism.

Does similar situation happen in the case of Lorentz symmetry? This
is a rather deep question since Lorentz symmetry is a fundamental
spacetime symmetry and has been incorporated into the two
cornerstones of current physics: general relativity and quantum
field theory. The possibility to think of Lorentz symmetry breaking
may be traced back to Dirac\cite{Dirac} through reintroducing aether
into the theory of electrodynamics in the early 1950s. There are
also other perspectives related to Lorentz symmetry violation
(LV)\cite{ELV}. It is first demonstrated by Kosteleck$\acute{y}$ and
Samuel that spontaneous Lorentz symmetry breaking may happen in
string field theory via unstable vacuum triggered by tachyon
field\cite{string}. After then Kosteleck$\acute{y}$ and Colladay
incorporated Lorentz symmetry violation into the effective theory
framework\cite{SME}, which is the so called standard model extension
(SME). In that work\cite{SME}, spontaneous Lorentz symmetry breaking
is triggered by nonzero vacuum expectation value (VEV) of a tensor
field in underlying theory, and these VEV of tensor fields are
incorporated with SM fields into all possible LV operators. In
addition to string motivated LV, other approaches of quantum gravity
also indicate some signatures of Lorentz violation. That include
spin-network calculation in loop quantum gravity\cite{loop}, foamy
structure of spacetime\cite{foam}, noncommutative quantum field
theory\cite{noncommutative}, and emergent gravity\cite{condense},
etc.. However, without a complete theory of quantum gravity, all
indications of LV above do not provide a firm and definite evidence
that Lorentz symmetry is indeed breaking, or in other words, why it
should not be an exact symmetry. However, we can take a positivism
viewpoint that we can rely on experiments to verify or put bound on
LV, as current experiments have already reached the sensitivity to
Planck mass suppression (e.g., for dimensionless couplings, the
sensitivity to Planck mass suppression means sensitivity to
$\frac{m_\mathrm{e}}{M_{\mathrm{Planck}}}\backsim10^{-23}$).

The purpose of this paper is to derive a set of modified dispersion
relations (MDR) in the framework of SME and explore their
consequence in the propagation properties of free particles. So we
first briefly review the basic principle of SME and its quantum
electrodynamics (QED) subset in sections 2 and 3. Then, by focusing
on the physical relevant LV couplings\cite{field redefinition} we
derive the MDR with CPT even and CPT odd LV couplings respectively,
together with referring their physical resemblance with other
distinct physical subjects in section 4. In section 5 the MDR is
applied to neutrino propagation and ultrahigh energy cosmic ray
(UHECR) problems separately. In that section we formulate a neutrino
oscillation mechanism in 2-spinor formalism and derive a much
stringent bound on the LV coupling involved in UHECR problem. In
section 6 we give a brief summary. The convention adopted in this
paper is $\eta_{00}=+1~~for~~\eta_{\mu\nu}$ and
$\epsilon_{0123}=+1~~for~~ \epsilon_{\rho\sigma\mu\nu}$.

%---------------------------------------------------
\section{Principle of SME  \label{sec:2}}
%---------------------------------------------------
The basic principle of standard model extension (SME) is that, SM is
regarded as a leading order Lagrangian in the low energy effective
field theory originating from a presumed existing fundamental
theory. While the other terms are treated as perturbation denoting
tiny departures from exact Lorentz symmetry. It is these
perturbation terms revealing the possible signature of physics
beyond SM. And the whole Lagrangian could be written as
\begin{equation}
\mathcal{L}=\mathcal{L}_{\mathrm{SM}}+\delta\mathcal{L},
\end{equation}
where $\delta\mathcal{L}$ is the Lagrangian denoting tiny LV
effects. Generally it has a form
\begin{eqnarray}
\delta\mathcal{L}\supset\frac{\lambda}{M^k}\left<T\right>\overline{\Psi}\Gamma(i\partial)^{k}\chi,
\end{eqnarray}
where the Lorentz indices of VEV of tensor field $\left<T\right>$
and partial differential operators $i\partial$ in (2) are
suppressed. These indices are matched so that they are contracted
exactly, which indicates that SME is apparent Lorentz covariant.
This is an explicit assumption of SME (i.e. LV terms are required to
be Lorentz covariant in their apparent Lorentz indices), and also a
direct consequence of the assumption of spontaneous Lorentz symmetry
breaking of an underlying Lorentz covariant theory, such as string
field theory\cite{string}. However, this covariance property of LV
operators should not be confused with particle Lorentz violation
they indicated. According to the work of Kostelecky and
Colladay\cite{SME}, observer Lorentz symmetry is nothing but the
equivalence relation of different coordinate choice, though
appropriate choice of coordinate system would largely simplify our
calculation and in some cases even would be helpful in the
interpretation of corresponding physical properties. While particle
Lorentz symmetry is a real symmetry concerning the properties of
identical particles (or localized fields) with different spin
orientation and momentum through particle rotation or boost
performed in a specified inertial frame. In ordinary theory, this is
just the symmetry classifying different species of identical
particles. While in LV theory, particle Lorentz transformation
leaves tensor VEV ($\left<T\right>$) unchanged, thus changes the
relation or interaction between SM fields with background tensor
fields, so new phenomena may arise. In this sense particle defined
as irreducible representation of Lorentz group is just a good
approximation if particle Lorentz symmetry is indeed violated.

Aside from the requirement of observer Lorentz invariance, other
restrictions may also help us to restrict or classify LV terms. We
could require the theory to be gauge invariant under a particular
gauge group transformations, e.g., gauge invariance under gauge
group $SU(3)_C\bigotimes SU(2)_L\bigotimes U(1)_Y $, and that is why
the theory is called SME. Dimensional counting may be used to
classify various LV operators. If restricted to dimension 3 or 4
terms, this is the minimal version of SME originally appeared
in\cite{SME}. We note that dimension 5 operators are also classified
recently\cite{Dimension}. Hermiticity and energy positivity are also
necessary to make the theory physically meaningful.

In addition to the above considerations, discrete symmetry
transformation can be applied on the LV operators to classify them
into CPT even and CPT odd classes. They form two special irreducible
representations of homogeneous Lorentz group respectively. Using the
convention of Coleman and Glashow\cite{High Energy}, the general
irreducible representation of homogeneous Lorentz group is marked by
$(A,B)$, where A,B are two angular momentum quantum numbers. So
$(1,1)$ is identified with CPT even operators and represents
traceless symmetric tensor of rank 2, while
$(\frac{1}{2},\frac{1}{2})$ is identified with CPT odd operators and
represents four-vector. According to the argument of Coleman and
Glashow\cite{High Energy}, the expectation value of $(A,A)$ operator
grows at large energy like $E^{2A}$. Thus CPT even operators
dominate at high energies.

%---------------------------------------------------
\section{QED Subset of SME  \label{sec:3}}
%---------------------------------------------------

In this section we present power-counting renormalizable QED subset
satisfying all the requirements discussed in the above section. We
can divide LV QED into pure photon part and fermion part. The
interaction between them is included through covariant derivatives
$\stackrel{\rightarrow}{\mathcal{D}^\nu}=\stackrel{\rightarrow}{\partial^\nu}+iq~A^\nu$.
For simplicity we confine ourselves to electrons though the equation
derived below can be applicable to more general fermions which are
not necessarily elementary particles. The LV QED Lagrangian is
\begin{equation}
\delta\mathcal{L}_{\mathrm{QED}}=\delta\mathcal{L}_{\mathrm{photon}}+\delta\mathcal{L}_{\mathrm{electron}},
\end{equation}
where
\begin{equation}
\delta\mathcal{L}_{\mathrm{photon}}\supset-\frac{1}{4}(k_F)_{\kappa\lambda\mu\nu}F^{\kappa\lambda}
         F^{\mu\nu}+\frac{1}{2}(k_{AF})_\kappa\epsilon^{\kappa\lambda\mu\nu}A_\lambda
         F_{\mu\nu},
\end{equation}
and
\begin{eqnarray}&&
\delta\mathcal{L}_{\mathrm{electron}}=\delta\mathcal{L}_{\mathrm{electron}}^{\mathrm{even}}+\delta\mathcal{L}_{\mathrm{electron}}^{\mathrm{odd}},\\
&&
\delta\mathcal{L}_{\mathrm{electron}}^{\mathrm{even}}\supset-\frac{1}{2}H_{\mu\nu}\overline{\psi}\sigma^{\mu\nu}\psi+\frac{i}{2}c_{\mu\nu}
         \overline{\psi}\gamma^\mu\stackrel{\leftrightarrow}{\mathcal{D}^\nu}\psi+\frac{i}{2}d_{\mu\nu}\overline{\psi}
         \gamma_5\gamma^\mu\stackrel{\leftrightarrow}{\mathcal{D}^\nu}\psi,
         \\ &&
\delta\mathcal{L}_{\mathrm{electron}}^{\mathrm{odd}}\supset-a_{\mu}\overline{\psi}\gamma^\mu{\psi}-b_{\mu}\overline{\psi}\gamma_5\gamma^\mu{\psi},
\end{eqnarray}
including those which are not directly deducible from terms
compatible with electroweak structure
\begin{equation}
\delta\mathcal{L}_{\mathrm{electron}}^{\mathrm{odd}}\supset\frac{i}{2}e^\nu\overline{\psi}\stackrel{\leftrightarrow}{\mathcal{D}}_\nu
         \psi-\frac{1}{2}f^\nu\overline{\psi}\gamma_5\stackrel{\leftrightarrow}{\mathcal{D}}_\nu\psi+\frac{1}{4}i
         g^{\lambda\mu\nu}\overline{\psi}\sigma_{\lambda\mu}\stackrel{\leftrightarrow}{\mathcal{D}}_\nu\psi.
\end{equation}

All the coupling coefficients $c$, $d$, $e$, $f$, $g$, $m_5$, $a$,
$b$, and $H$ above are real and constant parameters required by the
hermiticity of Lagrangian. They are related to VEV of tensor fields
in the underlying theory. However, not all of them are physically
observable, and some of them can be eliminated through field
redefinition. This is the result of the fact that there is a spinor
coordinate selection freedom, which implies that the mathematical
expression of Lorentz invariant Dirac Lagrangian is not uniquely
determined\cite{field redefinition}. There exists an equivalent
class of Dirac Lagrangian which are related by the fermion field
redefinition of the form
\begin{equation}
\Psi(x)\equiv[1+f(x,\partial)]\chi(x),
\end{equation}
where $f(x,\partial)$ represents a general $4\times4$ matrix
function of the coordinates and derivatives. For example, we can
choose $f(x,\partial)=+ia\cdot x$, or its finite form
$\Psi(x)=\exp[ia\cdot x]\chi(x)$ to reproduce
$-a_{\mu}\overline{\psi}\gamma^\mu{\psi}$ from the conventional
Dirac Lagrangian
\begin{equation}
 \mathcal{L}_{\mathrm{\mathrm{Dirac}}}=\frac{1}{2}i\overline{\psi}\gamma^\mu\stackrel{\leftrightarrow}{\partial}_\mu\psi-m\overline{\psi}
         \psi.
\end{equation}

%About the text below eq. (10): did I understand correctly that you
%ignore the issue of locality because the current work refers to the
%free fermions ? Is there still a way to proceed from the eqs. of
%motion (modified free Dirac eqs.) that you consider to an
%interacting theory ? Well, to some extent one might of course live
%with non-locality.

Thus some apparent (physical irrelevant) LV couplings can be
accounted for by field redefinition of conventional Dirac spinors,
and then can be absorbed into redefined fields through inverse
transformation. However, this field redefinition or field
transformation works effectively only in the absence of interaction
with other fields or interaction between spinor components due to
nonlocality problem, which were observed by Colladay and
McDonald\cite{field redefinition}. Fortunately, since in the
following we will focus on extended Dirac equation and its
consequence, disregard photon parts and set the covariant
derivatives $\stackrel{\leftrightarrow}{\mathcal{D}}_\nu$ into
partial derivatives $\stackrel{\leftrightarrow}{\partial}_\nu$, that
is to consider only free Dirac equations, so no such obstructions
will meet when performing field redefinition to remove some
couplings. After field redefinition, the simplified extended Dirac
Lagrangian which contains only physically relevant parameters (some
rearrangement is performed to include original Dirac Lagrangian to
form a compact and elegant form) is written as
\begin{eqnarray}
 \mathcal{L}_{\mathrm{electron}}=\frac{1}{2}i\overline{\psi}\widetilde{\Gamma}^\mu\stackrel{\leftrightarrow}{\partial}_\mu-\overline{\psi}
         \widetilde{M}\psi,
\end{eqnarray}
where
\begin{eqnarray} &&
    \widetilde{\Gamma}^\mu=\gamma^\mu+\widetilde{c}^{(\nu\mu)}\gamma_\nu+\widetilde{d}^{\nu\mu}\gamma_5\gamma_\nu+\frac{1}{2}
         \widetilde{g}^{\lambda\nu\mu}\sigma_{\lambda\nu},  \\ &&
    \widetilde{M}=m+\widetilde{b}_\mu\gamma_5\gamma^\mu+\frac{1}{2}\widetilde{H}_{\mu\nu}\sigma^{\mu\nu}.
\end{eqnarray}
All the coefficients in the above two equations (12) and (13) have
been redefined, thus do not have the symmetry properties as their
original ones (without tilde) in their corresponding Lorentz
indices. For details, see Colladay et al.\cite{field redefinition}.
For simplicity we omit the ``tilde" below and the reader should not
confuse them with the original ones.

Using the Euler-Lagrangian equation
\begin{equation}
\frac{\partial\mathcal{L}}{\partial\Psi^l}-\partial_\mu\frac{\partial\mathcal{L}}{\partial(\partial_\mu\Psi^l)}=0,
\end{equation}
we can obtain the extended Dirac equation below
\begin{equation}
[i(\gamma_\nu+c_{\mu\nu}\gamma^\mu+d_{\mu\nu}\gamma_5\gamma^\mu+\frac{g_{\lambda\mu\nu}}{2}\sigma^{\lambda\mu})\stackrel{\rightarrow}{\partial^\nu}-(m+b_\mu\gamma_5\gamma^\mu+\frac{1}{2}H_{\mu\nu}\sigma^{\mu\nu})]\Psi(x)=0.
\end{equation}
In subsequent section, we will discuss this equation in detail.

%---------------------------------------------------
\section{Derivation of Dispersion Relation of Extended Dirac
Equation  \label{sec:4}}
%---------------------------------------------------

In order to get a modified dispersion relation, we proceed with the
usual squaring procedure (which leads to Klein-Gordon equation when
we apply it to the usual Dirac equation) to see the consequence when
apply it to the extended Dirac equation (15). Multiplying (15) by
$[-i\Gamma^\rho\stackrel{\rightarrow}{\partial}_\rho-M]$, we get
equation
\begin{eqnarray}&&
[\Gamma^\rho\Gamma^\mu\stackrel{\rightarrow}{\partial}_\rho\stackrel{\rightarrow}{\partial}_\mu+i[\Gamma^\rho,M]\stackrel{\rightarrow}{\partial}_\rho+M^2]=0,
\end{eqnarray}
where
\begin{eqnarray}&&
\Gamma^\rho\Gamma^\mu\stackrel{\rightarrow}{\partial}_\rho\stackrel{\rightarrow}{\partial}_\mu=
\{G_{(\nu}^{~\rho)}G^{(\nu\mu)}-d_\nu^{~\rho}d^{\nu\mu}-2i\gamma_5\sigma_{\nu\sigma}d^{\nu\mu}G^{(\sigma\rho)}
 -\epsilon_{\lambda\nu\sigma\alpha}g^{\lambda\nu\mu}(G^{(\sigma\rho)}+\gamma_5d^{\sigma\rho})\gamma_5\gamma^\alpha
 \nonumber\\ &&~~~~~~~~~~ +\frac{1}{4}g^{\lambda\nu\mu}g^{\sigma\alpha\rho}[i\epsilon_{\lambda\nu\sigma\alpha}\gamma_5+(\eta_{\sigma\lambda}
 \eta_{\alpha\nu}-\eta_{\sigma\nu}\eta_{\alpha\lambda})]\}\stackrel{\rightarrow}{\partial}_\rho\stackrel{\rightarrow}{\partial_\mu},
  \\
&& M^2=m^2+m(2
b_\mu\gamma_5\gamma^\mu+H^{\rho\sigma}\sigma_{\rho\sigma})-b_\rho
b^\rho-\epsilon^{\mu\nu\sigma\alpha}H_{\mu\nu}b_\sigma\gamma_\alpha
\nonumber \\
&&~~~~~~~~~~~~~~+\frac{1}{4}H_{\mu\nu}H_{\rho\sigma}[i\epsilon^{\mu\nu\rho\sigma}\gamma_5+(\eta^{\rho\mu}\eta^{\sigma\nu}-\eta^{\rho\nu}\eta^{\sigma\mu})],
\\ &&
  [\Gamma^\rho,M]=g^{\lambda\eta\rho}H^{\mu\nu}(\eta_{\eta\nu}\eta_{\lambda\mu}-\eta_{\eta\mu}\eta_{\lambda\nu})
  +ig^{\lambda\nu\rho}b^\mu\gamma_5(\eta_{\mu\nu}\gamma_\lambda-\eta_{\mu\lambda}\gamma_\nu)
  -\nonumber\\ &&~~~~~~~~~~ 2G^{\mu\rho}b_\mu\gamma_5+iH^{\mu\nu}(G^{\sigma\rho}+\gamma_5d^{\sigma\rho})(\eta_{\mu\sigma}\gamma_\nu-\eta_{\nu\sigma}\gamma_\mu)
  -2id^{\nu\rho}b^\mu\sigma_{\mu\nu}.
\end{eqnarray}
In equation (17) and (19) we use the definition
\begin{eqnarray}[\eta^{\sigma\rho}+c^{(\sigma\rho)}]=G^{(\sigma\rho)}.\end{eqnarray}
Note that, with the symmetry property of $c^{(\sigma\rho)}$, we have
\begin{eqnarray}G_{(\nu}^{~\rho)}G^{(\nu\mu)}=\eta^{\rho\mu}+c_\nu^{~\rho}c^{(\nu\mu)}+2c^{(\rho\mu)}.\end{eqnarray}
Obviously this squared extended Dirac equation with 2 classes of
undetermined LV parameters (
$c_{(\mu\nu)},~d_{\mu\nu},~H_{\mu\nu};~~b_\mu,~g_{\lambda\nu\mu}$)
is too complicated to be diagonalized in spinor space by continuing
the same procedure.  However, since their CPT properties are
distinct, we can discuss them separately. But we should keep in mind
that this is just for convenience. There is no fundamental reason to
forbid the appearance of the CPT odd operators unless one imposes
CPT symmetry as a custodial symmetry survived after the Lorentz
symmetry breaking. Note that Lorentz invariance is just one part of
sufficient conditions to the proof of CPT theorem in the local field
theory, not a necessary one. On the other hand, CPT violation
conclusively leads to Lorentz violation in local field theory, a
theorem proved by Greenberg\cite{Greenberg}. So we could have a
Lorentz violating theory with only CPT even operators, in which CPT
odd operators are all ruled out by CPT invariance. While in the
theory with CPT odd operators (this theory would be automatically
Lorentz violating, as guaranteed by Greenberg's theorem ), CPT even
Lorentz violating operators would be induced via loops, so they must
be much smaller in the naive analysis with the assumption of tree
level disappearance of CPT even Lorentz violating operators, thus
can be neglected at the tree level calculation, which will be the
case of next subsection.

%---------------------------------------------------
%\subsubsection{CPT Odd   \label{sec:A}}
\subsection{CPT Odd   \label{sec:A}}
%---------------------------------------------------
At first we write the field equation involving only CPT odd LV
couplings
\begin{equation}
[i\gamma\cdot{\partial}+\frac{i}{2}g_{\lambda\mu\nu}\sigma^{\lambda\mu}{\partial^\nu}-(m+\gamma_5b\cdot\gamma)]\Psi(x)=0,
\end{equation}
then by multiplying (22) on the left with
$[-(i\gamma\cdot{\partial}+\frac{i}{2}g_{\lambda\mu\nu}\sigma^{\lambda\mu}{\partial^\nu}-\gamma_5b\cdot\gamma)-m]$,
we obtain a quadratic equation
\begin{eqnarray}&&
\{[\eta_{\nu\rho}+\frac{1}{4}(i\epsilon^{\lambda\mu\alpha\beta}\gamma_5+(\eta^{\lambda\alpha}\eta^{\mu\beta}-\eta^{\lambda\beta}\eta^{\mu\alpha}))g_{\lambda\mu\nu}g_{\alpha\beta\rho}]\partial^\nu\partial^\rho+b^2+m^2
\nonumber \\
&&+2\gamma_5\sigma_{\rho\nu}b^\rho\partial^\nu-i\epsilon^{\lambda\rho\mu\alpha}b_\rho
g_{\lambda\mu\nu}\gamma_\alpha\partial^\nu\}\Psi(x)=0.
\end{eqnarray}
This equation can be rearranged by putting all diagonal terms (in
spinor space) on one side, while leaving nondiagonal ones on the
other side, that is
\begin{eqnarray}&&
\{[\partial^2+\frac{1}{2}g_{\lambda\mu\nu}g^{\lambda\mu}_{~~~\rho}\partial^\nu\partial^\rho+b^2+m^2]+\nonumber
\\&&
[\frac{i}{4}\epsilon^{\lambda\mu\alpha\beta}g_{\lambda\mu\nu}(\gamma_5~g_{\alpha\beta\rho}\partial^\rho+4\gamma_\alpha~b_\beta)\partial^\nu+2\gamma_5\sigma_{\rho\nu}b^\rho\partial^\nu]\}\Psi(x)=0.
\end{eqnarray}
Proceeding with the same squaring procedure, in principle we can get
an 8th order differential equation without the appearance of
$\Gamma$ structure matrices. However, since this routine is too
tedious and makes physics obscure, we do not follow this way, rather
we concentrate on the equation (22) itself.  As noted in\cite{SME},
$g_{\lambda\mu\nu}$ may arise from interaction among fermion
constituents for a composed fermion, thus is expected to be
suppressed further more than other LV couplings, so we simply drop
it in (22) and get
\begin{equation}
[i\gamma\cdot\partial-(m+\gamma_5b\cdot\gamma)]\Psi(x)=0.
\end{equation}
It would be easy to get a quartic order differential equation
\begin{equation}
[({\partial^2}+b^2+m^2)^2-4((b\cdot(i\partial))^2+b^2\partial^2)]\Psi(x)=0
\end{equation}
from equation (25). By using the Ansatz
\begin{equation}
\Psi(x)\equiv\phi(p)\exp[-ip\cdot x],
\end{equation}
we finally get a modified dispersion relation
\begin{equation}
[(p^2-b^2-m^2)^2+4b^2p^2-4(b\cdot p)^2]=0.
\end{equation}
Note that, equation (28) is noninvariant under interchange
$p\rightarrow-p$, however it is invariant under simultaneous
interchange $p\rightarrow-p$ and $b\rightarrow-b$.  This is the
common feature of CPT odd LV operators, which indicates a helicity
dependence of energy levels. Taking into account of
$g_{\lambda\mu\nu}$ just implies a further splitting of energy
degeneracy.

Before closing this subsection, we observe that even without LV,
gravity can induce an equation of the same form as (25) except that,
the constant vector $b_\mu$ is replaced by a spacetime dependent
vector $B_a$, whose meaning will become clear later. This implies
that gravitational field provides a practical global Lorentz
symmetry breaking source. The above observation will be manifested
by the derivation of the covariant Dirac equation
\begin{equation}
\mathcal{L}_{\mathrm{spin-gravity}}=\sqrt{-g}[i\overline{\psi}\gamma^a\stackrel{\rightarrow}{\mathcal{D}}_a\psi-m\overline{\psi}
       \psi],
\end{equation}
where the covariant derivative is
$\stackrel{\rightarrow}{\mathcal{D}}_a=\stackrel{\rightarrow}{\partial}_a-\frac{1}{4}w_{bca}\sigma^{bc}$.
Thus
\begin{eqnarray}&&
\mathcal{L}_{\mathrm{spin-gravity}}=\sqrt{-g}[i\overline{\psi}\gamma\cdot\partial\psi-m\overline{\psi}\psi]+[\frac{\sqrt{-g}}{4}\overline{\psi}\gamma^a\sigma^{bc}\psi
w_{bca}]\nonumber \\
&&
~~~~~~~~~~~~~~~~=\mathcal{L}_{\mathrm{free}}+\mathcal{L}_{\mathrm{int}},
\end{eqnarray}
where
$\mathcal{L}_{\mathrm{int}}=\frac{\sqrt{-g}}{4}\overline{\psi}\gamma^a\sigma^{bc}\psi
w_{bca}$ can be shown equal to
\begin{eqnarray}&&
\mathcal{L}_{\mathrm{int}}=\mathcal{L}_{\mathrm{VI}}+\mathcal{L}_{\mathrm{AI}}\nonumber \\
&&
~~~~~~=\frac{i\sqrt{-g}}{2}\overline{\psi}\eta^{a[b}\gamma^{c]}\sigma^{bc}\psi
w_{[bc]a}+\frac{\sqrt{-g}}{4}\overline{\psi}\gamma_5\gamma_d\psi\epsilon^{abcd}w_{[bc]a},
\end{eqnarray}
where $w_{[bc]a}=w_{bca}$ and $\mathcal{L}_{\mathrm{VI}}$ is an
antihermitian term, thus vanishes automatically by the hermiticiy
requirement of the theory. While $\mathcal{L}_{\mathrm{AI}}$ can be
shown equal to $\sqrt{-g}\overline{\psi}\gamma_5\gamma_d\psi B^d$,
where
\begin{equation}
B^d\equiv\frac{1}{4}\epsilon^{abcd}w_{[bc]a}=\frac{1}{4}\epsilon^{abcd}e_{b\rho}(\partial_a~e_c^{~\rho}+\Gamma^\rho_{~\mu\nu}e_a^{~\nu}e_c^{~\mu}).
\end{equation}
Thus (29) could be rewritten in the form of
\begin{eqnarray}&&
\mathcal{L}_{\mathrm{spin-gravity}}=\mathcal{L}_{\mathrm{free}}+\mathcal{L}_{\mathrm{AI}}\nonumber \\
&&
~~~~~~~~~~~~~~~~=\det(e)\overline{\psi}[(i\gamma\cdot\partial-m)+\gamma_5\gamma_d
B^d]\psi.
\end{eqnarray}
By identifying $B_a$ with $-b_a$, we see that the covariant Dirac
equation
\begin{eqnarray}
[(i\gamma\cdot\partial-m)+\gamma_5\gamma_d B^d]\psi=0
\end{eqnarray}
is of the same form as (25).

As observed by Mohanty, Prasanna, and Lambiase\cite{inflation
leptongenesis}, the spin-gravity coupling can induce leptongenesis
in the presence of lepton number violation interactions, thus may
help to resolve the net baryon asymmetry problem through the so
called electroweak sphaleron process. Since equation (34) concerning
spin-gravity coupling and equation (25) concerning LV vector
coupling are of the same form, a non-vanishing CPT odd LV coupling,
$b_\mu$, may also be a candidate in demonstrating these effects and
thus provide a possible solution to asymmetry problem.
%Moreover£¬ since the covariant Dirac equation can be solved exactly
%to first order in the metric deviation as given by
%%Papini\cite{interferometry}, we expect that the same issue may be
%also applicable to (4.10) since $b_\mu$ has been strictly
%constrained by experiments to be tiny. Thus we may solve (4.10)
%formally in exactly the same way as solving
%(4.19)\cite{interferometry}.
So the formal similarity of the two equations suggests that $b_\mu$
and $-B_a$ may mimic the effects each other produced, thus
experimental searches of the two may be complementary. In other
words, experimental searches for curvature-spin coupling may also
provide signals for LV bounds on $b_\mu$, and vice versa. However,
there is a significant difference between the two. First, $b_\mu$ in
LV case is treated as constant background field, while $-B_a$
generated by curvature couplings is spacetime dependent though in
some cases can be treated as semi-classical background. Second,
$b_\mu$ is a CPT odd LV coupling treated as an unaltered constant
under CPT transformation, while $-B_a$ is generated from
gravitational sources and should transform in the same way as the
ordinary matter field under CPT, thus makes the corresponding
operator CPT invariant, which could be easily seen from the fact
that gravitational interaction respect CPT symmetry. Further more,
though $-B_a$ breaks global Lorentz symmetry, it respects local
Lorentz symmetry automatically in an appropriate free fall inertial
frame, thus it is actually a local Lorentz invariant term. The last
point is that, $-B_a$ is a universal coupling reflecting a curved
spacetime effect provided we insist on the equivalence principle to
be still hold in quantum-gravity interplay region, while there is no
good reason to regard $b_\mu$ as universal. So experimentally we
could distinguish the two with different physical meanings by the
effects produced by nonvanishing $\delta b_\mu=b^i_\mu-b^j_\mu$
(where $i$ and $j$ refer to different flavors).

%---------------------------------------------------
%\subsubsection{CPT even  \label{sec:B}}
\subsection{CPT Even  \label{sec:B}}
%---------------------------------------------------
Next we turn to CPT even LV corrections to conventional Dirac
equation and derive the corresponding dispersion relations.

For completeness, we rewrite the CPT even Dirac Lagrangian below
\begin{equation}
\mathcal{L}^{\mathrm{even}}_{\mathrm{electron}}=\frac{i}{2}\overline{\psi}(\eta_{\mu\nu}+c_{\mu\nu}-d_{\mu\nu}\gamma_5)\gamma^\mu\stackrel{\leftrightarrow}{\partial^\nu}\psi-\overline{\psi}
         (m+\frac{1}{2M}\stackrel{\thicksim}{H}_{\mu\nu}\sigma^{\mu\nu})\psi,
\end{equation}
where we replace $H_{\mu\nu}$ by
$\frac{1}{M}\stackrel{\thicksim}{H}_{\mu\nu}$, i.e.
\begin{equation}
\stackrel{\thicksim}{H}_{\mu\nu}\equiv H_{\mu\nu}\times M.
\end{equation}
The meaning of this replacement will be clear later. Then we write
down the corresponding field equation by setting $g_{\lambda\mu\nu}$
and $b_\mu$ equal to zero in (15), that is
\begin{equation}
[i(\gamma_\nu+c_{\mu\nu}\gamma^\mu+d_{\mu\nu}\gamma_5\gamma^\mu)\stackrel{\rightarrow}{\partial^\nu}-(m+\frac{1}{2}H_{\mu\nu}\sigma^{\mu\nu})]\Psi(x)=0.
\end{equation}
Multiplying (37) on the left with
$[i(\gamma_\nu+c_{\mu\nu}\gamma^\mu+d_{\mu\nu}\gamma_5\gamma^\mu)\stackrel{\rightarrow}{\partial^\nu}-\frac{1}{2}H_{\mu\nu}\sigma^{\mu\nu}+m]$,
we get the following equation
\begin{eqnarray}&&
\{-[G_{\rho\mu}G^\rho_{~\nu}-d_{\rho\mu}d^\rho_{~\nu}-2i\sigma_{\rho\sigma}d^\rho_{~\mu}[\gamma_5G^\sigma_{~\nu}+d^\sigma_{~\nu}]]\partial^\mu\partial^\nu
 +i\epsilon_{\alpha\rho\sigma\beta}[\gamma_5G^\alpha_{~\nu}+d^\alpha_{~\nu}]\gamma^\beta
 H^{\rho\sigma}\partial^\nu\nonumber \\
&&~~~~~~~~~~~~~+\frac{1}{4}[i\epsilon_{\alpha\rho\sigma\beta}\gamma_5H^{\alpha\rho}H^{\sigma\beta}+2H^2]-m^2\}\Psi(x)=0.
\end{eqnarray}
In deriving this equation, we use the anti-commutation relations
\begin{eqnarray}&&
\{\gamma_\nu,
\sigma_{\rho\sigma}\}=\{\gamma_\nu,\frac{i}{2}[\gamma_\rho\gamma_\sigma-\gamma_\sigma\gamma_\rho]\}=-2\epsilon_{\nu\rho\sigma\alpha}\gamma_5\gamma^\alpha,
\\ &&
\{\sigma_{\rho\sigma},\sigma_{\mu\nu}\}=2i\epsilon_{\rho\sigma\mu\nu}\gamma_5+2(\eta_{\rho\mu}\eta_{\sigma\nu}-\eta_{\rho\nu}\eta_{\sigma\mu}),
\end{eqnarray}
which can be proven by direct calculation. Note that, for
aesthetical consideration, we have retained the term
$-2i\sigma_{\rho\sigma}d^\rho_{~\mu}d^\sigma_{~\nu}\partial^\mu\partial^\nu$
in
$-2i\sigma_{\rho\sigma}d^\rho_{~\mu}[\gamma_5(c+\eta)^\sigma_{~\nu}+d^\sigma_{~\nu}]\partial^\mu\partial^\nu$,
which vanishes automatically for antisymmetric properties of
$\sigma_{\rho\sigma}$. In (38) we use $G_{\mu\nu}$ defined in (20)
instead of $c_{\mu\nu}$, and this definition is triggered by the
observation that the Minkowvsky metric $\eta_{\mu\nu}$ is always
followed by $c_{\mu\nu}$.
%In application of the modified dispersion
%relation to neutrino problems, we will see that $c_{\mu\nu}$ does
%play part of the role of fluctuation of metric in a nontrivial way
%in some case.

Eq. (38) shows that, it is hard to be diagonalized in 4-spinor space
by the squaring procedure we used before. This is due to the
entanglement between nondiagonal terms involving $d_{\mu\nu}$ and
$H_{\mu\nu}$. So we can derive the modified dispersion relation by
assuming $d_{\mu\nu}=0$ or $H_{\mu\nu}=0$ respectively.

When $H_{\mu\nu}=0$ in (38), it leads to equation
\begin{eqnarray}&&
\{[(\eta_{\mu\nu}+c_{\rho\mu}c^\rho_{~\nu}+2c_{\mu\nu}-d_{\rho\mu}d^\rho_{~\nu})\partial^\mu\partial^\nu+m^2][(\eta_{\alpha\beta}+c_{\sigma\alpha}c^\sigma_{~\beta}+2c_{\alpha\beta}-d_{\sigma\alpha}d^\sigma_{~\beta})\partial^\alpha\partial^\beta+m^2]+\nonumber\\
&&
4[d_{\rho\alpha}d^\rho_{~\mu}(c+\eta)^\sigma_{~\nu}(c+\eta)_{\sigma\beta}-d^\rho_{~\mu}(c+\eta)_{\rho\beta}d^\gamma_{~\alpha}(c+\eta)_{\gamma\nu}]\partial^\mu\partial^\nu\partial^\alpha\partial^\beta\}\Psi(x)=0.
\end{eqnarray}
This remains to be a complicated equation. By using Ansatz (27) we
can get a quartic-order dispersion relation
\begin{eqnarray}&&
[(G_{\rho\mu}G^\rho_{~\nu}-d_{\rho\mu}d^\rho_{~\nu})p^\mu~p^\nu-m^2][(G_{\sigma\alpha}G^\sigma_{~\beta}-d_{\sigma\alpha}d^\sigma_{~\beta})p^\alpha
~p^\beta-m^2]\nonumber \\&&
+4[d_{\rho\alpha}d^\rho_{~\mu}G^\sigma_{~\nu}G_{\sigma\beta}-d^\rho_{~\mu}G_{\rho\beta}d^\gamma_{~\alpha}G_{\gamma\nu}]p^\mu~p^\nu~p^\alpha~p^\beta=0,
\end{eqnarray}
and this is still a complicated equation. We can analyze the role of
$d_{\mu\nu}$ and $c_{\mu\nu}$ separately in (42) by setting the
opposite term equal to zero respectively.

Setting $d_{\mu\nu}=0$ corresponds to
\begin{eqnarray}&&
[(\eta_{\mu\nu}+c_{\rho\mu}c^\rho_{~\nu}+2c_{\mu\nu})p^\mu~p^\nu-m^2]=(G_{\rho\mu}G^\rho_{~\nu}p^\mu~p^\nu-m^2)\nonumber \\
&& ~~~~~\equiv~(\widetilde{G}_{\mu\nu}p^\mu~p^\nu-m^2)=0,
\end{eqnarray}
where at the last step we define $\widetilde{G}_{\mu\nu}\equiv
G_{\rho\mu}G^\rho_{~\nu}$. This definition makes (43) looking like a
formally relativistic dispersion relation, except with Minkowvsky
metric $\eta_{\mu\nu}$ replaced by $\widetilde{G}_{\mu\nu}$. From
(43) and the definition (20) we see that $c_{\mu\nu}$ behaves like a
fluctuation of metric.%  previous referred.

While setting $c_{\mu\nu}=0$ in (42) corresponds to
\begin{equation}
[(p^2+m^2+d_{\rho\mu}d^\rho_{~\nu}p^\mu~p^\nu)^2-4(m^2p^2+(d_{\rho\mu}p^\rho~p^\mu)^2)]=0,
\end{equation}
which is a quartic order equation, but could be solved formally by
using triangle parametrization with
\begin{eqnarray}&&
X^2\equiv~4m^2p^2, \nonumber \\
&& Y^2\equiv~4(d_{\rho\mu}p^\rho~p^\mu)^2, \nonumber \\
&& Z^2\equiv~(p^2+m^2+d_{\rho\mu}d^\rho_{~\nu}p^\mu~p^\nu)^2.
\end{eqnarray}
Thus $X^2+Y^2=Z^2$, which is just the identity of (44), and set
\begin{eqnarray}
Y=Z\sin[\theta],~~~~~X=Z\cos[\theta],
\end{eqnarray} with $\theta\lll1$ since $d_{\mu\nu}$
constrained by experiment must be small.

When $d_{\mu\nu}=0$ in (38), it leads to equation
\begin{eqnarray}&&
\{[(\eta_{\mu\nu}+c_{\rho\mu}c^\rho_{~\nu}+2c_{\mu\nu})\partial^\mu\partial^\nu-\frac{1}{2}H^2+m^2]^2+\nonumber\\&&
[\epsilon_{\alpha\rho\sigma\beta}\epsilon_{\gamma\zeta\eta\delta}H^{\alpha\rho}H^{\gamma\zeta}(\frac{1}{16}H^{\sigma\beta}H^{\eta\delta}-\eta^{\beta\delta}G^\sigma_{~\nu}G^\eta_{~\xi}\partial^\nu\partial^\xi)]\}\Psi(x)=0,
\end{eqnarray}
where  $H^2=H^{\zeta\eta}H_{\zeta\eta}$.

When replacing $i\partial^\mu\rightarrow~p^\mu$ in (47), we can
solve the corresponding quartic order equation in the quadratic form
\begin{eqnarray}&&
p^2+(c_{\rho\mu}c^\rho_{~\nu}+2c_{\mu\nu})p^\mu~p^\nu+\frac{1}{2}H^2-m^2=0,
\\ &&
\epsilon_{\alpha\rho\sigma\beta}\epsilon_{\gamma\zeta\eta\delta}H^{\alpha\rho}H^{\gamma\zeta}(\frac{1}{16}H^{\sigma\beta}H^{\eta\delta}+\eta^{\beta\delta}G^\sigma_{~\nu}G^\eta_{~\xi}{p^\nu}{p^\xi})=0,
\end{eqnarray}
where (49) should be interpreted as a constraint equation on $H$. It
is possible to separate (48) from (47) only in the case of the
assumption that (49) is semi-positive defined. This assumption is
satisfied definitely when the vector defined below
\begin{eqnarray}&&
  J_\beta\equiv\frac{1}{2}\epsilon_{\alpha\rho\sigma\beta}H^{\alpha\rho}G^\sigma_{~\nu}p^\nu
\end{eqnarray} is a timelike or lightlike vector, which can be seen from the expression (49). In
equation (49) we see $H$ appears with a totally antisymmetric tensor
$\epsilon_{\gamma\zeta\eta\delta}$, its appearance reminisces us the
anomaly expression in the presence of gauge field, that is
$\mathcal{A}(x)=-\frac{1}{16\pi^2}\epsilon_{\mu\nu\rho\sigma}F^{\mu\nu}_\alpha(x)F^{\rho\sigma}_\beta(x)tr[t_\alpha%
     t_\beta t]$\cite{QFT}. So we guess the physical effects produced by $\widetilde{H}^{\mu\nu}$ defined in (36) (rather than
$H^{\mu\nu}$) resemble that of electromagnetic field. This is indeed
confirmed by the operator $-\frac{1}{2M}\overline{\psi}
         \stackrel{\thicksim}{H}_{\mu\nu}\sigma^{\mu\nu}\psi$ in the Lagrangian (35),
which is nothing but the ``Pauli term"\cite{QFT} appeared in the
Lagrangian form. This term could give an additional contribution to
fermion magnetic moment (anomalous magnetic moment), thus could be
constrained by muon ``g-2" experiment.

%of course the term quoted from [16] (although it was not new there)
%is a gauge term, hence a bit different from eq. (59). Also the
%anomalous magnetic moment of the muon is a matter of gauge
%interactions. The latter is not explained well by theory even
%without LV terms, essentially because the non-perturbative QCD
%contribution is not known precisely.

%---------------------------------------------------
\section{Applications  \label{sec:5}}
%---------------------------------------------------

Since we have already derived a set of modified dispersion relations
induced by various LV couplings, we can see what novel physical
consequences these relations can lead to. It is well known that
$E^2=\overrightarrow{p}^2+m^2$ is a fundamental equation in
conventional physics, so modification of this equation is expected to have a  %profound
wide impact on high energy physics at which LV effects are expected
to be less suppressed than at low energies. Indeed, introducing even
minuscule LV would lead to processes conventionally forbidden at
high energies, or accumulating unexpected observable effects when
particles propagate through cosmological distance, or even lead to
some processes allowed at intermediate energy range while forbidden
at higher and lower energies\cite{High Energy}. For example,
radiative muon decay $\mu\rightarrow~e+\gamma$\cite{High Energy},
neutron stability\cite{High Energy} at ultrahigh energies, and
vacuum dispersion and birefringence\cite{Carroll-2001}, vacuum
photon splitting\cite{vacuum}, and photon decay\cite{Jacobson}, etc.

However, for the dispersion relations we derived, there are two
points to be clarified. Firstly, there is a preferred frame in which
each relation has a most simplified form. So when we use this form
of relation, we implicitly presume that a preferred inertial frame
has been chosen. Secondly, as we previously commented, the LV
couplings involved are redefined couplings, while this redefinition
only works properly in the absence of interactions. So, strictly
speaking, these relations are only applicable to propagating
problems where particles involved could be considered as free
fermions, though no obstruction would meet in deriving a dispersion
relation from the interaction equation, where partial derivatives
are replaced by covariant derivatives and radiative corrections are
include to calculate a complete propagator. Then, since CPT-odd
equation resembles the covariant Dirac equation with spin-gravity
coupling involved, which had been extensively discussed
elsewhere\cite{inflation leptongenesis}, and CPT-odd operator grows
with energy increase much slower than CPT-even one, which has been
discussed in the end of section II, we focus our attention on
CPT-even couplings and discuss the implication of the corresponding
MDR in neutrino and ultrahigh energy cosmic ray problems.

%---------------------------------------------------
\subsection{Neutrino Oscillation \label{sec:Ab}}
%---------------------------------------------------

Neutrino oscillation might be the only definite signal indicating
physics beyond SM, and has been extensively discussed in the
literature\cite{neutrino}. While most of them focus on neutrino with
mass nondegenerate scenario and use Dirac equation as a starting
point, which makes the assumption of neutrino classification
unclear. In our derivation, we assume neutrinos to be Majorana
spinors from the beginning, thus it is suited to be described in a
2-spinor formalism. As this assumption indicates, we should first
reduce the Dirac equation into a 2-spinor form. Beginning with (37)
by ignoring $H_{\mu\nu}$ for simplicity, we rewrite the equation as
\begin{equation}
[i(\gamma_\nu+c_{\mu\nu}\gamma^\mu+d_{\mu\nu}\gamma_5\gamma^\mu)\stackrel{\rightarrow}{\partial^\nu}-m]\Psi(x)=0.
\end{equation}
For convenience, we redefine these LV couplings in a manifest V-A
form resembling the V-A theory, which is a low energy effective
field theory of electroweak theory. The definition is
\begin{eqnarray}&&
g^L_{\mu\nu}\equiv(c-d)_{\mu\nu},\quad
g^R_{\mu\nu}\equiv(c+d)_{\mu\nu}.
\end{eqnarray}
With this definition, equation (51) could be written as
\begin{equation}
[i(\gamma_\nu+g^L_{\mu\nu}\frac{1-\gamma_5}{2}\gamma^\mu+g^R_{\mu\nu}\frac{1+\gamma_5}{2}\gamma^\mu)\stackrel{\rightarrow}{\partial^\nu}-m]\Psi(x)=0.
\end{equation}
Using the projection operator $\frac{1\pm\gamma_5}{2}$ and the
definition of Majorana spinor
$\Psi\equiv\Psi^c=\mathcal{C}\overline{\Psi}^T$, we can derive from
(53) the corresponding equation
\begin{equation}
i(\sigma_\nu+c_{\mu\nu}\sigma^\mu)\stackrel{\rightarrow}{\partial^\nu}\phi-im\sigma^2\phi^\star=0
\end{equation}
satisfied by Majorana 2-spinors, where
$\sigma^\mu\equiv(-1,\stackrel{\rightarrow}{\sigma})$ and  $\phi$ is
the reduced wave function. For details, see Appendix. Rewrite (54)
in the form of Schr\"{o}dinger equation
\begin{eqnarray}&&
i\frac{\partial}{\partial~t}\phi=\frac{1}{i}\stackrel{\rightarrow}{\sigma}\cdot\stackrel{\rightarrow}{\nabla}\phi-ic_{\mu\nu}\sigma^\mu\stackrel{\rightarrow}{\partial^\nu}\phi,
\end{eqnarray}
where we already assumed neutrino to be massless fermion, as our
derivation of neutrino oscillation will not be based on mass
nondegenerate scenario. Note that $\phi$ is a simple notation of a
column of 2-spinors, and in our case, we consider only two flavors
as an illustration, so $\phi\equiv\left(
                                              \begin{array}{c}
                                               |\nu_\mu\rangle \\
                                               |\nu_\tau\rangle \\
                                              \end{array}
                                            \right)$,
and the corresponding Hamiltonian
$\widehat{H}=\frac{1}{i}\stackrel{\rightarrow}{\sigma}\cdot\stackrel{\rightarrow}{\nabla}-ic_{\mu\nu}\sigma^\mu\partial^\nu$
should be regarded as a $2\times2$ matrix operator in flavor space.
Using the Ansatz $\phi(x)\equiv\phi(p)\exp[-ip\cdot x]$, we can
write the Hamiltonian in momentum space as
\begin{eqnarray}
\widehat{H}=\left(
                                         \begin{array}{cc}
                                         \stackrel{\rightarrow}{\sigma}\cdot\stackrel{\rightarrow}{p}-(c_{\mu\nu})_{11}\sigma^\mu~p^\nu & -(c_{\mu\nu})_{12}\sigma^\mu~p^\nu \\
                                           -(c_{\mu\nu})_{21}\sigma^\mu~p^\nu &\stackrel{\rightarrow}{\sigma}\cdot\stackrel{\rightarrow}{p}-(c_{\mu\nu})_{22}\sigma^\mu~p^\nu \\
                                         \end{array}
                                       \right).
\end{eqnarray}
We can diagonalize this Hamiltonian by a rotation matrix $R$. Choose
a specific reference frame in which rotation invariance still holds,
then we can assume
$(c_{ij})_{\alpha\beta}=k_{\alpha\beta}\delta_{ij}$ (where $i$ and
$j$ run over 1 to 3, $\alpha$ and $\beta$ run over 1 to 2), and that
all the other terms are zero. Since Lorentz violation is stringently
restricted to be tiny, $k_{\alpha\beta}\ll1$, we can simply drop the
diagonal terms of $({c_{\mu\nu}})_{\alpha\beta}$ in (56), and assume
$k_{12}=k_{21}=k$. Then we can get the corresponding eigenvalues of
Hamiltonian (56) as
\begin{eqnarray}
 \lambda_1=1+k, \quad
 \lambda_2=1-k,
\end{eqnarray}
and the corresponding rotation matrix
\begin{eqnarray}
R\equiv\left(\begin{array}{cc}
            \cos[\theta] & -\sin[\theta] \\
            \sin[\theta] & \cos[\theta] \\
          \end{array} \right)=\frac{1}{\sqrt{2}}\left(
                        \begin{array}{cc}
                          1 & -1 \\
                          1 & 1 \\
                        \end{array}
                      \right),
\end{eqnarray} which just corresponds to set the rotation angle in (58)
as $\theta=\frac{\pi}{4}$. Then we can get the relation between
energy eigenvector $R\phi$ with $\phi$
\begin{eqnarray}
 \left(
  \begin{array}{c}
    |\nu_\mu\rangle \\
    |\nu_\tau\rangle \\
  \end{array}
\right)=\left(
          \begin{array}{cc}
            \cos[\theta] & \sin[\theta] \\
            -\sin[\theta] & \cos[\theta] \\
          \end{array}
        \right)
        \left(
                      \begin{array}{c}
                        |\nu\rangle_1 \\
                         |\nu\rangle_2 \\
                      \end{array}
                    \right).
\end{eqnarray}
For a muon-type neutrino emitted, the neutrino state after evolving
through a time interval $t$ is determined by
\begin{eqnarray}&&
|\nu(t)\rangle=(\sin[\theta]\exp[-i\delta E\cdot
t]|\nu\rangle_2+\cos[\theta]|\nu\rangle_1)\exp[-iE_1t]\nonumber
\\&&
~~~~~=[\frac{1}{2}\sin[2\theta](\exp[-i\delta E\cdot
t]-1)|\nu_\tau\rangle\nonumber \\
&&~~~~~~~~~~+(1+\sin[\theta]^2(\exp[-i\delta E\cdot
t]-1))|\nu_\mu\rangle]\exp[-iE_1t],
\end{eqnarray}
so the flavor transition probability is given by
\begin{eqnarray}&&
P_{\mu\rightarrow~\tau}(t)=|\langle\nu_\tau|\nu(t)\rangle|^2=\sin^2[2\theta]\sin^2[\frac{\delta
E\cdot t}{2}]\nonumber
\\&&~~~~~~~~~~~
\simeq~\sin^2[2\theta]\sin^2[\frac{|\overrightarrow{p}|2k\cdot
t}{2}],
\end{eqnarray}
where at the last step we used (57) and $\delta
E=E_2-E_1=-2k|\overrightarrow{p}| $. Including the mass terms in
(55) and (56) just complicates our formula without any principled
difficulty. Note that the presence of nondiagonal tensor couplings
($(c_{\mu\nu})_{\alpha\beta}\neq0$ for $\alpha\neq\beta$) in flavor
space is essential for this Lorentz violation induced neutrino
oscillation scenario. Though this is a rather simple model to
illustrate neutrino oscillation caused by tiny Lorentz violation, we
can still gain some insight by comparing it with experiments. The
MINOS experiments reported their oscillation fit results with
$\sin^2[2\theta_{23}]>0.84$ with $90\%$ confidence level and
$\bigtriangleup
m^2_{23}=2.38^{+0.20}_{-0.16}\times10^{-3}\mathrm{eV}^2$ with $68\%$
confidence level\cite{MINOS} and they analyzed the data with the
same two flavor $\nu_\mu\to \nu_\tau$ oscillation assumption. We
find that the oscillation angle $\theta=\frac{\pi}{4}$ is consistent
with MINOS results, even very close to it. From the squared mass
difference ($\sim10^{-3}\mathrm{eV}^2$) and the robust bound from
cosmology on the sum of neutrino mass\cite{Cosmosn} ($0.5-1.0$~eV),
we can estimate the neutrino mass to be around 0.1~eV order. Then
Lorentz violation coupling would contribute to an effective mass
term as can be seen from (54) when neutrino is significantly
energetic for $k\cdot |\overrightarrow{p}|\sim
m_{\mathrm{neutrino}}$. In the MINOS neutrino experiment, the peak
in neutrino energy spectrum is around 3~GeV~(low-energy beam) to
7.8~GeV~(high-energy beam)\cite{MINOS}, so we can give a rather
rough bound on the size of Lorentz violation coupling
$k\sim\frac{m_{\mathrm{neutrino}}}{|\overrightarrow{p}|}\sim\frac{0.1\mathrm{eV}}{1\mathrm{GeV}}\sim10^{-10}$,
if Lorentz violation contribution to neutrino oscillation in MINOS
experiment is comparable to the non-degenerate mass contribution. In
principle, this bound could be restricted to more stringent accuracy
of order $k\sim10^{-22}\sim\frac{\bigtriangleup m^2_{23}}{E^2}$ by
dimensional analysis. Since this estimate of the size of Lorentz
violation coupling is just the inverse of $\gamma(=\frac{E}{m})$
factor of the high energy neutrino and depends on the assumption of
comparably contribution of Lorentz violation, to obtain more
accurate estimate of the order of this LV coefficient we need to
take into account the mass effect (i.e., (54) is used) and give a
more reasonable weight on Lorentz violation contribution by using
experiment data (e.g., $\bigtriangleup m^2_{23}$) directly or even
by matching the whole energy spectrum. We noticed that some more
comprehensive work\cite{LV Oscillation} have already been done in
the three flavor case which involves the whole renormalizable LV
operators ($c$, $d$, $a$, $b$, $e$, $f$, $g$, and $H$), though this
was done under some reasonable perturbative expansion (since Lorentz
violation correction would be tiny) in order to get the required
effective Hamiltonian which controls neutrino propagation effects.
In their first paper\cite{LV Oscillation}, a general framework in
the LV induced neutrino oscillation was given and some definite
signals in experimental searches for Lorentz violation in neutrino
sector were classified and examined. Their subsequent works focused
on particular models where the number of nonzero LV parameters were
reduced significantly\cite{LV Oscillation}. In this sense, our work
is just a illustration or toy-model, however, its simplicity makes
the oscillation mechanism induced by tiny Lorentz violation more
obviously and the assumption of the neutrino property (Majorana
neutrino) more apparently. Further it could be generalized to more
practical model (3-flavors) directly by taking into account mass
terms since the pure Lorentz violation (i.e., massless neutrino
case) model may not be a practical solution to globally fit all
neutrino oscillation data from solar, reactor and atmosphere
neutrino experiments\cite{Challenge}. However, whether the
generalized form could accommodate with the experiment data or not
still lacks checking.
%The latter (energy spectrum matching) is a more complete analysis
%and may give a more reliable estimate on Lorentz violation but
%beyond the scope of this paper.

Some remarks should be said about the transition probability (61)
which is proportional to neutrino energy, instead of inverse
proportional to it as in the case of mass nondegenerate scenario.
This property is the common feature of all nonstandard neutrino
oscillation scenarios and reminisces us the neutrino oscillation
induced by equivalent principle violation\cite{EP}. The formula of
which is exactly the same as (61), except replacing $2k$ with
$\frac{h_{00}}{2}$, where $h_{00}=-2\phi=\frac{2 G M\alpha}{R}$ is
the 00-component of metric fluctuation, and $\alpha$ is the post
Newtonian parameter (in general relativity, $\alpha=1$ and is
universal). This similarity is not an accident, since Lorentz
violation is assumed as an remnant of quantum gravity, and in
quantum region there is some indication that equivalence principle
is violated. We guess that the equivalence principle violation may
indicate Lorentz violation at least locally, as the equivalence
principle ensures the existence of local inertial frames, which is
the premise of local Lorentz transformation.  Actually, Lorentz
violation must be followed by equivalence principle violation, an
issue recently clarified in\cite{EPV}. Furthermore, we can see from
(52) that nonvanishing $d_{\mu\nu}$ gives rise to different
couplings to left and right handed Dirac fermions, so it may induce
observable effects in energy splitting between different helicities.

%---------------------------------------------------
\subsection{Ultrahigh Energy Cosmic Ray \label{sec:Bb}}
%---------------------------------------------------

Ultrahigh energy cosmic ray (UHECR) provides a natural source of
high energy particles, with energies up to $10^{19}~$eV, much higher
than that of energetic particles generated by man-made accelerator.
But the energy of UHECR reached earth can not be much higher than
that, as it is predicted to be terminated at around
$5\times10^{19}~$eV for the energy lose in the collision of UHECR
particle with CMB photons by Greisen, Zatsepin and
Kuz'min\cite{GZK}, which is known as the GZK cutoff. Similar
situation happens in the collision of TeV $\gamma$ ray with infrared
photons. However, this prediction has not yet been confirmed by
experiments. AGASA, Fly's Eye both claimed that they observed events
with energies nearer or above this cutoff\cite{absence}, while
HiRes\cite{confirm1} and Pierre Auger\cite{confirm2} experiments
recently claimed the observation of the cutoff. This unsettled
problem has stimulated many attempts to resolve it, including active
galactic nuclei (which is favored by Pierre Auger experiment),
primary flux of magnetic monopoles, ``Z-boson bursts" produced by
collision of UHE neutrino with relic neutrino nearby, pseudo-complex
extension of standard field theory\cite{LIGZK}, etc.. Of course, LV
also provides a possible candidate to extend or entirely rule out
this cutoff\cite{High Energy}. In this paper, we follow the general
analysis of Coleman and Glashow\cite{High Energy} and show that the
LV coupling in modified dispersion relation (43) could be
constrained either from the absence or the presence of GZK cutoff.

We take the common assumption that the identity of UHECR are
protons, and analyze the pion-nucleon resonance formation reaction
$P~+~\gamma(\mathrm{CMB})~\to~\bigtriangleup(1232)$, which is the
dominant contribution to GZK cutoff. This reaction is possible if
and only if
$E_0\geq~E_{\mathrm{min}}(\overrightarrow{\mathrm{P}}_0)$, where
$E_0$ is the total energy of initial particles and
$E_{\mathrm{min}}(\overrightarrow{\mathrm{P}}_0)$ denotes the
minimum total energy of the final products, whose total momentum is
equal to initial total momentum $\overrightarrow{\mathrm{P}}_0$,
which is implicitly assumed from energy momentum conservation. So
the reaction is kinematically allowed by the condition
\begin{eqnarray}
\omega~+~E_p\geq~E_\triangle,
\end{eqnarray}
where $\omega$ is the energy of CMB photon, $E_\mathrm{p}$ and
$E_\triangle$ are the energies of proton and $\triangle$ resonance,
with the subscripts denoting proton and $\triangle$ resonance
respectively. We rewrite dispersion relation (43) in a relativistic
form
\begin{eqnarray}
E_a^2=\overrightarrow{P}_a^2c_a^2+m_a^2c_a^4,
\end{eqnarray}
where $c_a$ is the maximal attainable velocity for the $a$th
particle defined in\cite{High Energy}. In the case we considered, it
just requires the definition below
\begin{eqnarray}
c_a\equiv\frac{1}{1+c_{00}^a},
\end{eqnarray}
where $c_{00}^a$ denotes the $00$ component of $c_{\mu\nu}$ for the
$a$th fermion, and other components of which are assumed to be zero.
We rewrite (62) in terms of $E_p$ and $\omega$,
\begin{eqnarray}
\omega~+~E_p\geq~\sqrt{(|\overrightarrow{P}_p|-\omega)^2c_\triangle^2+m_\triangle^2c_\triangle^4}.
\end{eqnarray}
For threshold reaction, the initial proton momentum
$\overrightarrow{P}_p$ is collinear with that of $\triangle$
resonance while anticollinear with that of CMB photon, which has
already been used in (65), i.e., the substitution of
$\overrightarrow{P}_\bigtriangleup=\overrightarrow{P}_p-\omega$.
Squaring (65), we have
\begin{eqnarray}&&
E_p^2(1+\frac{c_\triangle}{c_p})(1-\frac{c_\triangle}{c_p})+2\omega(E_p+|\overrightarrow{P}_p|c_\triangle^2)+
\omega^2(1-c_\triangle^2)\nonumber \\
&&~~~~~~~~~~~~~~~~~~~~~~~~~~~~~~~~~~+(m_p^2-m_\triangle^2\frac{c_\triangle^2}{c_p^2})c_p^2c_\triangle^2\geq0.
\end{eqnarray}
For ultrahigh energy proton: $E_p\sim|\overrightarrow{P}_p|$. As LV
coupling should be much smaller than 1, we take the approximation
that $c_\triangle\sim1$ and $1+\frac{c_\triangle}{c_p}\sim2$.
Substituting this approximation into (66) leads to a quadratic order
inequality of $E_p$
\begin{eqnarray}&&
2E_p^2(1-\frac{c_\triangle}{c_p})+4\omega~E_p+ K\geq0,\end{eqnarray}
where
\begin{eqnarray}K\equiv(m_p^2-m_\triangle^2\frac{c_\triangle^2}{c_p^2})c_p^2c_\triangle^2.
\end{eqnarray}
Thus the threshold energy that the reaction kinematically allows is
the small positive root of (67) when equality is hold. With the
assumption
\begin{eqnarray}
1-\frac{c_\triangle}{c_p}>0,
\end{eqnarray}
and quadratic order equation (67) (with equality hold), we obtain
two roots with opposite signs. One is negative but with larger
absolute value, the other is positive which gives the threshold
energy, i.e., pion-nucleon resonance formation reaction is
kinematically allowed only for energy above this positive value. On
the other hand, with the assumption
\begin{eqnarray}
1-\frac{c_\triangle}{c_p}<0,
\end{eqnarray}
we obtain two positive values. The small one is the threshold
energy, while the larger one is the terminating energy of this
reaction, which means that the formation reaction is kinematically
allowed in an intermediate energy band. This is a striking feature
of Lorentz violation corrections to the familiar particle reaction
process previously referred and was systematically discussed
in\cite{High Energy}, so GZK problem in this case provides one
concrete illustration of the analysis in\cite{High Energy}. However,
both cases give the same threshold formula, that is
\begin{eqnarray}&&
E_\mathrm{p}=\frac{\omega\sqrt{1-1/2\frac{K}{\omega^2}(1-\frac{c_\triangle}{c_p})}-\omega}{1-\frac{c_\triangle}{c_p}}\nonumber
\\ &&
~~~~~~\simeq-\frac{K}{4\omega}-\frac{K^2}{32\omega^3}(1-\frac{c_\triangle}{c_p})+\ldots,
\end{eqnarray}
where the first term at the last step is the conventional threshold
energy in the absence of LV, and the other terms are tiny LV
corrections. Substituting
$E_{\mathrm{thre}}=-\frac{K}{4\omega}=5\times10^{19}$~eV\cite{GZK}
%with $c_a$ in it replaced by 1
and the experimentally observed threshold energy% energy breaking in UHECR energy spectrum
$E_p=5.6\times10^{19}$~eV\cite{confirm1} into (71), we can deduce
the bound
\begin{eqnarray}
 1-\frac{c_\triangle}{c_p}=-\frac{2w(E_p-E_{\mathrm{thre}})}{E_{\mathrm{thre}}^2}.
\end{eqnarray}
Since $\triangle$ resonance is not a spin-1/2 fermion, we simply
assume $c_\triangle=1$, and substituting (64) into (72) yields the
bound on LV coupling $c_{00}^p\backsim~10^{-25}$, which is more
stringent than that derived in\cite{High Energy} by two orders of
magnitude. This is consistent with our expectation since we adopt
the data confirming GZK cutoff. Note that this bound is of
importance only in the sense of order of magnitude, since it cannot
be fixed firmly from the location of GZK cutoff alone, many other
effects could compensate for some amount of Lorentz violation, for
example, the uncertainty of source distribution.
%could be further
%restricted if we could obtain a more accurate threshold energy
%$E_\mathrm{p}$ in UHECR energy spectrum experimentally.
However, the bound we obtained is strict enough and has already
reached the Planck mass suppression sensitivity$10^{-23}$, which
indicates that Lorentz violation in dimension 4 operators (for
proton-LV tensor coupling) is indeed too minuscule to be detected.

%Actually I thought that HiRes essentially claims that their
%observations are in full agreement with the GZK cutoff, so that the
%enumerator in eq. (72) would vanish. But in any case, just now
%people rather refer to the new results by the Pierre Auger
%Collaboration. According to their interpretation they agree with the
%GZK cutoff as well, E_thre can only be exceeded if the source is
%nearby. But actually the range of the AGN sources that they suspect
%is even a bit shorter than necessary for the proton attenuation
%length, hence a number of authors now suspect that the primary rays
%in these cases were actually light nuclei but not just protons, see
%e.g. arXiv:0801.0227 and arXiv:0804.2466.

%---------------------------------------------------
\section{Summary  \label{sec:6}}
%---------------------------------------------------

Searching for Lorentz violation (LV) experimentally or theoretically
has received much attention in recent years. As QED has been tested
to a marvelous accuracy, it is expected to be an ideal research area
to probe the minuscule trace of LV both in theory and in experiment.
In this paper we studied various modified dispersion relations (MDR)
derived from extended QED by the assumption that a particular set of
LV couplings is nonzero. This is a reasonable assumption since if a
fundamental theory really violates Lorentz symmetry, the
corresponding tensor coupling in low energy effective theory should
contain less parameters than what we presented here. On focusing
these LV couplings in MDR or extended Dirac equation, we observe the
similarity of some LV couplings with the spin-gravity couplings or
metric couplings in covariant Dirac equation. This resemblance may
indicate a deep physical relevance of these LV couplings with that
in the quantum-gravity interplay region, since Lorentz violation is
assumed as a remnant signal of quantum gravity.

In addition to that, this similarity has also been observed in the
application to the neutrino sector. We also derived an oscillation
formalism by explicitly assuming neutrino as Majorana spinor. We
found that the nondiagonal terms of LV couplings in flavor space
involved in neutrino sector could explain neutrino oscillation even
in the massless case. Though this possibility has been
comprehensively discussed by several authors\cite{LV
Oscillation}\cite{Challenge} and even have already been partly
tested in some experiment such as LSND\cite{LSND}, it is the first
time, as far as we known, to derive the oscillation by explicitly
assuming neutrino as Majorana spinor in 2-spinor formalism, thus our
model could be viewed as a toy-model simply demonstrating neutrino
oscillation induced by Lorentz violation. We also made a rough
estimate on the size of LV couplings involved as $k\sim10^{-10}$. We
note that the LV couplings for different species of fermions
involved in specific problems are defined independently since we
cannot calculate them from an underlying concrete model displaying
this LV effective Lagrangian as the low energy limit after
spontaneous Lorentz symmetry breaking. In principle this oscillation
formalism can be generalized to the 3 flavor case by taking into
account the mass effect. Then by comparing it with the neutrino
energy spectrum obtained experimentally rather than only with the
mixing angle and mass square difference, we expect that more
accurate bounds on LV couplings could be obtained in neutrino
sector.

By application of MDR to GZK problem, we derived a much stringent
bound on the order of magnitude of LV parameters of protons
($c_{00}^p\backsim~10^{-25}$) from the recent observation in
HiRes\cite{confirm1} and Pierre Auger experiments\cite{confirm2} and
we note that more bounds could be obtained on neutrons by taking
account of some exotic process (such as proton vacuum Cerenkov
radiation) in the analysis of the propagation of ultrahigh energy
cosmic rays when Lorentz violation is present\cite{Proton}. There
are also many stringent bounds on the magnitude of various LV
parameters\cite{bounds} up to date. We observe that most of these
bounds are actually the bounds on the difference of LV parameters to
different species of particles, as in the case of nonuniversal
gravity coupling induced neutrino oscillation\cite{EP} or the
maximal attainable velocity analysis developed in\cite{High Energy}.
They just indicate (from the opposite side) that the difference of
Lorentz violation tensor couplings to different species of particles
is rather small, in other words, they strongly suggest that the
tensor field triggering Lorentz violation in the underlying theory
couples to the standard model field universally, at least for
dimension 3/4 operators. So whether Lorentz symmetry is just a
perfectly good approximate symmetry is still an open question.

%---------------------------------------------------
\section*{Acknowledgments}
%---------------------------------------------------

We are grateful to Wolfgang Bietenholz, Bin Chen and Shi-Min Yang
for helpful discussions. This work is partially supported by
National Natural Science Foundation of China (Nos.~10721063,
10575003, 10528510), by the Key Grant Project of Chinese Ministry of
Education (No.~305001), and by the Research Fund for the Doctoral
Program of Higher Education (China).

%---------------------------------------------------

\appendix
%\section*{Appendix}

\section{}

%---------------------------------------------------

To get an equation describing Majorana 2-spinor in (54), we begin
with (53) describing Dirac 4-spinor. Actually, we could use 2-spinor
or 4-spinor formalism to describe either Dirac or Majorana fermions.
However, for a Majorana fermion the independent degrees of freedom
are 2, thus not all of its 4 components in 4-spinor formalism are
independent, while for a Dirac fermion, it needs at least 2
different 2-spinors. So for a Majorana fermion, it is adequate to be
described by 2-spinor formalism, while 4-spinor formalism is
suitable to describe a Dirac fermion. In order to get the suited
equation describing Majorana 2-spinor from (53), we need to use the
projection operator $\frac{1\pm\gamma_5}{2}$ and the Majorana spinor
definition $\Psi\equiv\Psi^c=\mathcal{C}\overline{\Psi}^T$ to
project equation (53) from $4\times4$ matrix space to the
irreducible $2\times2$ subspace.

First, we give the projected wave function and $\Gamma$ matrices
below
\begin{eqnarray}&&
\Psi^L\equiv\frac{1-\gamma_5}{2}\Psi,~~\Psi^R\equiv\frac{1+\gamma_5}{2}\Psi;
\\  &&
\gamma^L\equiv\frac{1-\gamma_5}{2}\gamma,~~\gamma^R\equiv\frac{1+\gamma_5}{2}\gamma.
\end{eqnarray}
Using these definitions to rewrite (53) in the form
\begin{eqnarray}&&
[i(\gamma_\nu^R+g^R_{\mu\nu}\gamma^{R\mu})\stackrel{\rightarrow}{\partial^\nu}\Psi^L+i(\gamma_\nu^L+g^L_{\mu\nu}\gamma^{L\mu})\stackrel{\rightarrow}{\partial^\nu}\Psi^R-m(\Psi^R+\Psi^L)]=0,
\end{eqnarray} and multiplying (A.3) from the left with $\gamma_5$, we
have
\begin{eqnarray}&&
[i(\gamma_\nu^R+g^R_{\mu\nu}\gamma^{R\mu})\stackrel{\rightarrow}{\partial^\nu}\Psi^L-i(\gamma_\nu^L+g^L_{\mu\nu}\gamma^{L\mu})\stackrel{\rightarrow}{\partial^\nu}\Psi^R-m(\Psi^R-\Psi^L)]=0.
\end{eqnarray}
From (A.3) and (A.4) we can get two independent equations
\begin{eqnarray}&&
[i(\gamma_\nu^R+g^R_{\mu\nu}\gamma^{R\mu})\stackrel{\rightarrow}{\partial^\nu}\Psi^L-m\Psi^R]=0,\nonumber
\\ &&
[i(\gamma_\nu^L+g^L_{\mu\nu}\gamma^{L\mu})\stackrel{\rightarrow}{\partial^\nu}\Psi^R-m\Psi^L]=0
\end{eqnarray}for left
handed and right handed fermions respectively, with the mass term
mixing each other. The two above equations are not independent from
each other for a Majorana fermion. We can take complex conjugate
operation on (53) and charge conjugate operation
$\Psi^c=\mathcal{C}\overline{\Psi}^T$ on wavefunction to get a
charge conjugate equation of the same fermion field
\begin{equation}
[i(\gamma_\nu+g^R_{\mu\nu}\frac{1-\gamma_5}{2}\gamma^\mu+g^L_{\mu\nu}\frac{1+\gamma_5}{2}\gamma^\mu)\stackrel{\rightarrow}{\partial^\nu}-m]\Psi^c(x)=0,
\end{equation}
which is just (53) with $g^L_{\mu\nu}$ and $g^R_{\mu\nu}$
interchanged. Since the charge conjugate field for a Majorana field
is just the original field multiplied with a phase factor (see
(A.7)), the field equation satisfied for the charge conjugate field
should be the same, thus impose the condition
$g^L_{\mu\nu}=g^R_{\mu\nu}$, i.e.,
$g^L_{\mu\nu}=g^R_{\mu\nu}=c_{\mu\nu}$. Then we can use the same
procedure above in getting (A.5) to get a set of corresponding
equations for (A.6), which is nothing but the same equations of
(A.5). Note that for Majorana field, equation (A.5) contains
actually just one independent equation, the lower one is the
equivalent form of the upper one. Using the condition satisfied by
Majorana fermion below
\begin{equation}
\Psi(x)=\eta\Psi^c(x),
\end{equation}
where $\eta$ is a phase factor (for simplicity, we take it equal to
1), we can get the relation
\begin{equation}
\Psi^R(x)=\frac{1+\gamma_5}{2}\Psi^c(x)=-\gamma^0\mathcal{C}\Psi^{L\star}(x).
\end{equation}
Then substituting (A.8) into (A.5), we get the equation below
\begin{eqnarray}&&
i(\eta_\nu+c_{\mu\nu}\eta^\mu)\stackrel{\rightarrow}{\partial^\nu}\phi+m\phi^\star=0,
\end{eqnarray} where
\begin{eqnarray}&&
\eta^\mu=\mathcal{C}^{-1}\gamma^0\gamma^{R\mu} \\
&&  \phi(x)=\Psi^L(x).
\end{eqnarray}

Since (A.9) is expressed in $4\times4$ matrix space, thus could be
reducible. We can find an irreducible representation of $\eta^\mu$
in $2\times2$ matrix space, that is setting
$\eta^\mu=i\sigma^2\sigma^\mu$, where $\sigma^\mu\equiv(-1,
\overrightarrow{\sigma})$. So in 2-spinor irreducible space, (A.9)
becomes
\begin{equation}
i(\sigma_\nu+c_{\mu\nu}\sigma^\mu)\stackrel{\rightarrow}{\partial^\nu}\phi-im\sigma^2\phi^\star=0,
\end{equation}which is just (54).

We can derive this equation satisfied by Majorana fermion from a
more manifest way in displaying its Majorana feature, i.e.,
rewriting the modified Dirac Lagrangian in 2-component formalism and
using the neutrality condition to drop the coupling terms between
the two fermions. We first give the 2-component formalism of
modified Dirac Lagrangian, which is obtained by expressing
\begin{eqnarray}&& \psi_D=\left(\begin{array}{c}
                                                                   \chi \\
                                                                   \overline{\eta}
                                                                 \end{array}\right)
,\quad \gamma^{\mu}=\left(
                 \begin{array}{cc}
                   0 & \sigma^{\mu} \\
                   \overline{\sigma}^{\mu} & 0 \\
                 \end{array}
               \right)
\end{eqnarray} in the Lagrangian (35), where $\sigma^\mu\equiv(-1,
\overrightarrow{\sigma})$, $\overline{\sigma}^{\mu}\equiv(-1,
-\overrightarrow{\sigma})$. The resulting Lagrangian (only contains
Lorentz violating couplings) is
\begin{eqnarray}&&
\mathcal{L}_{\mathrm{Dirac}}=-i\eta(c+d)_{\mu\nu}\sigma^{\mu}\partial^{\nu}\overline{\eta}-i\overline{\chi}(c-d)_{\mu\nu}\overline{\sigma}^{\mu}\partial^{\nu}\chi
\nonumber \\
&&~~~~~~~~~~~+iH_{\mu\nu}(\eta\sigma^{\mu\nu}\chi+\overline{\chi}\overline{\sigma}^{\mu\nu}\overline{\eta}).
\end{eqnarray}
Note $\sigma^{\mu\nu}$ appears here is defined to be
$\sigma^{\mu\nu}=\frac{1}{4}(\sigma^\mu\overline{\sigma}^\nu-\sigma^\nu\overline{\sigma}^\mu)$,
where
$\overline{\sigma}^{\mu\nu}=\frac{1}{4}(\overline{\sigma}^\mu\sigma^\nu-\overline{\sigma}^\nu\sigma^\mu)$,
rather then that appeared in the text, which could be relabeled by
$\Sigma^{\mu\nu}=\frac{i}{2}[\gamma^\mu, \gamma^\nu]=2i\left(
                                                         \begin{array}{cc}
                                                          \sigma^{\mu\nu} & 0 \\
                                                           0 & \overline{\sigma}^{\mu\nu} \\
                                                         \end{array}
                                                       \right)
$. For any Dirac spinor $\psi_D$, we can construct two corresponding
Majorana spinors as \begin{eqnarray}&&
\psi_{M1}=\frac{1}{\sqrt{2}}(\psi_D+\psi_D^c),\quad
\psi_{M2}=\frac{-i}{\sqrt{2}}(\psi_D-\psi_D^c),
\end{eqnarray}where $\psi_D^c=\mathcal{C}\overline{\psi_D}^T$ is
just the charge conjugate fermion field. Using this equation we can
decompose the two Weyl-spinors as two decoupled Majorana spinors in
the absence of additional inter-couplings (e.g., LV-couplings) by
the deduced equation below\begin{eqnarray}&&
\chi=\frac{1}{\sqrt{2}}(\psi_1+i\psi_2),\quad
\eta=\frac{1}{\sqrt{2}}(\psi_1-i\psi_2),
\end{eqnarray}where $\Psi_{Mi}=\left(
                   \begin{array}{c}
                     \psi_i \\
                     \overline{\psi_i} \\
                   \end{array}
                 \right), \quad i=1,2.
$

Then the corresponding Lagrangian (not the full one (35)) is
\begin{eqnarray}&&
\mathcal{L}_{\mathrm{Dirac}}=-ic_{\mu\nu}\sum_{i=1}^{2}(\psi_i\sigma^{\mu}\partial^{\nu}\overline{\psi_i})+d_{\mu\nu}(\psi_1\sigma^{\mu}\partial^{\nu}\overline{\psi_2}-\psi_2\sigma^{\mu}\partial^{\nu}\overline{\psi_1})
\nonumber \\
&&~~~~~~~~~~~
+H_{\mu\nu}(\psi_2\sigma^{\mu\nu}\psi_1+\overline{\psi_2}\overline{\sigma}^{\mu\nu}\overline{\psi_1}),
\end{eqnarray}
where we have used the fact $\psi\sigma^{\mu\nu}\psi=0$. So we can
explicitly see the couplings between two Weyl-spinors arising from
$d_{\mu\nu}$ and $H_{\mu\nu}$. In the Majorana theory, we can simply
drop them. Thus the corresponding full Majorana-Lagrangian in
2-component theory is
\begin{eqnarray}&&
\mathcal{L}^2_{\mathrm{Majorana}}=i(\eta_{\mu\nu}+c_{\mu\nu})\partial^{\nu}\psi\sigma^{\mu}\overline{\psi}+\frac{m}{2}(\psi\psi+\overline{\psi}\overline{\psi}),
\end{eqnarray} and the corresponding 4-component form is
\begin{eqnarray}&&
\mathcal{L}^4_{\mathrm{Majorana}}=\frac{i}{2}\overline{\Psi}(\gamma^\mu+c^{\nu\mu}\gamma_\nu)\stackrel{\leftrightarrow}{\partial}_\mu\Psi-\frac{m}{2}\overline{\Psi}\Psi.
\end{eqnarray}
From the Majorana Lagrangian (A.18), we can deduce an equation
\begin{eqnarray}
i(\eta_{\mu\nu}+c_{\mu\nu})\sigma^{\mu}\partial^{\nu}\overline{\psi}-m\psi=0.
\end{eqnarray}
From the definition $\Psi^c_M=\lambda\Psi_M$, where $\Psi=\left(
                                                            \begin{array}{c}
                                                              \psi \\
                                                              \overline{\psi} \\
                                                            \end{array}
                                                          \right),
$ and the convention $\lambda=1$ being chosen (which could be
confirmed from equation (A.15)), we can deduce
\begin{eqnarray}
\psi=i\sigma^2\overline{\psi}^\star.
\end{eqnarray}
Substituting this equation to (A.20), we again obtain equation
(A.12).
%---------------------------------------------------
% Bibliography
%---------------------------------------------------

\end{document}